# Peak Nothing: Recent Trends in Mineral Resource Production

James R. Rustad[1,2]
[1]Corning Inc., Corning, NY, 14831, [2]University of California, Davis, Davis, CA 95616

**Abstract**-The production histories of seventeen raw materials are analyzed with the logistic model. Although many of these resources have exhibited logistic behavior in the past, they now show exponential or super-exponential growth. In most cases, the transition has occurred in the last ten to twenty years.

Underpinning the recent focus on sustainability is the assumption that, for many raw materials, resource depletion is imminent *(1)*. The degree of concern has been compounded by increasing public awareness of the 'Hubbert's Peak' approach to oil production *(2)*. In this model, the consequences of resource depletion are apparent not near the end of production, but immediately after the peak of production where falling supplies meet rising demand. In such a context, the era of depletion is widely thought to have already begun. Applied initially to oil, the logistic Hubbert approach can shed light on the reserves of other elements.

The logistic equation, as applied to resource production, can be written:

$$d\ln Q(t)/dt = s*[1-(Q(t)/Q_{tot})] \qquad (1)$$

where *t* is time, $Q(t)$ is the cumulative production, $Q_{tot}$ is the total amount of the resource available, and *s* is a parameter which can be interpreted as the initial rate of production at *t*=0. For a nearly infinite resource ($Q(t)/Q_{tot} << 1$), this equation reduces to the familiar equation for exponential growth. An important idea in Equation (1) is to convert the yearly production to a fractional production rate, by dividing the yearly production by the cumulative production. If we let *P=dQ/dt* Equation 1 can be rewritten:

$$P/Q = s - \frac{s}{Q_{tot}} Q$$

Yearly production histories for a number of non-renewable resources have been compiled by the United States Geological Survey *(3)*. These data $P(n\Delta t)=(\Delta Q(n\Delta t)/\Delta t)$ (where *n* is an integer $\Delta t$ is one year) may analyzed within the logistic model by plotting cumulative production $Q(n\Delta t)$ on the x-axis and the fractional growth rate $P(n\Delta t)/Q(n\Delta t)$ on the y-axis *(4)*. In this picture, cumulative production has replaced time as the independent variable, and yearly production has been normalized by cumulative production.

If the use of a given resource is increasing exponentially, this type of plot gives a flat line, indicating that the fractional growth rate *P/Q* is a constant percent per year, independent of cumulative production. If the production is logistic, *P/Q* starts at some initial rate at *Q*=0 and decreases linearly with cumulative production until the resource is exhausted. At this point, cumulative production equals the total amount of the resource available, $Q_{tot}$, and *P/Q* goes to zero.

Figure 1 gives *P/Q* versus *Q* plots made from world production histories of a variety of non-renewable resources taken from *(3)*. Data for world oil production from 1930-2005 were obtained from *(4)*. In each case, the cumulative production has been normalized by dividing by the current cumulative production. For the data in *(4)*, pre-historical production is taken to be zero. This assumption has negligible effect on cumulative production values greater than ten percent of the current cumulative production. Thus the independent variable ranges between 0.1 and 1.

As shown in Figure 1, world oil production passes through an early super-exponential phase of growth up to about 1970, followed by a first logistic trend from about 1970 to the early 1980s pointed at an apparent $Q_{tot}$ of approximately 700-900 gigabarrels. This phase is followed by a second logistic trend from the early 1980's to the present, pointing to a second $Q_{tot}$ of approximately 2000 gigabarrels.

In many cases, the plots for other non-renewable resources resemble the one for oil. An initial trend is established pointing towards some apparent $Q_{tot}^*$ which is then replaced by a different trend, pointing toward a new $Q_{tot} \gg Q_{tot}^*$. Plots made in 1990 for bismuth and phosphorous; 1995 for platinum, germanium, cobalt and tellurium; and 2000 for zirconium, manganese, and molybdenum would have resulted in apparent $Q_{tot}$ values nearly equal to current cumulative production. Subsequently, production of these resources has transitioned to exponential or super-exponential growth. Several resources often popularly perceived as exhibiting "peak" or logistic behavior, such as the rare earth elements, lithium, and helium show no evidence for a finite $Q_{tot}$ at any point in their production histories. Because of the tendency of established trends to change abruptly, the logistic model cannot be used to estimate a convincing $Q_{tot}$ for any of the resources displayed in Figure 1.

The implications of Figure 1 are not necessarily optimistic. The fact that production of a given resource is growing exponentially does not imply that the resource is inexhaustible. The main implication of Figure 1 is that while many strategic resources appear to exhibit logistic trends during their production histories, these trends have changed to exponential or even super exponential growth over the last 10-20 years.

**References and Notes**

1. J. L. Sznopek, W. M. Brown *Materials Flow and Sustainability* United States Geological Survey Fact Sheet FS-068-98 (1998).

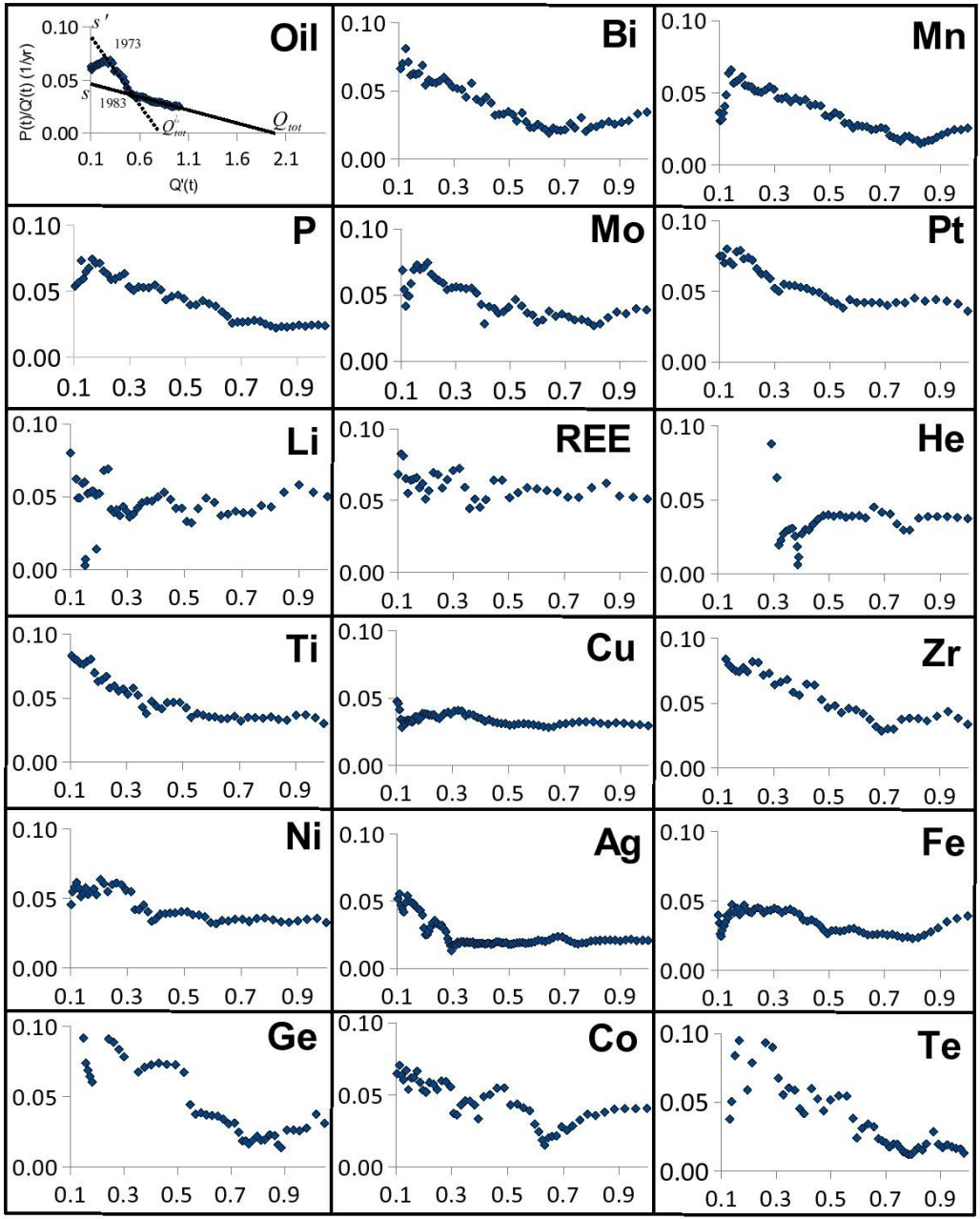

Figure 1. Plots of $P(t)/Q'(t)$ (in units of yr$^{-1}$) versus $Q'(t)$ for the production histories of seventeen of the resources compiled in *(3)*. $P(t)$ is the production per year, and $Q'(t)$ is the dimensionless cumulative production (cumulative production divided by the current cumulative production). REE stands for rare earth elements which are not differentiated in *(3)*.